\documentclass[9pt,english,sigconf,nonacm]{acmart}
\PassOptionsToPackage{noend,linesnumbered,ruled}{algorithm2e}
\usepackage[T1]{fontenc}
\usepackage[utf8]{inputenc}
\usepackage{parskip}
\usepackage{refstyle}
\usepackage{mathrsfs}
\usepackage{mathtools}
\usepackage{algorithm2e}
\usepackage{amsthm}
\usepackage{graphicx}

\makeatletter
\numberwithin{equation}{section}
\numberwithin{figure}{section}

\usepackage{tikz}
\usetikzlibrary{decorations.pathreplacing}

\let\citet\citep

\makeatother

\theoremstyle{plain}
\newtheorem{thm}{\protect\theoremname}[section]
\theoremstyle{definition}
\newtheorem{defn}[thm]{\protect\definitionname}
\theoremstyle{plain}
\newtheorem{fact}[thm]{\protect\factname}
\theoremstyle{remark}
\newtheorem{notation}[thm]{\protect\notationname}
\theoremstyle{definition}

\usepackage{babel}
\newref{thm}{
        name      = \RSthmtxt,
        names     = \RSthmstxt,
        Name      = \RSThmtxt,
        Names     = \RSThmstxt,
        rngtxt    = \RSrngtxt,
        lsttwotxt = \RSlsttwotxt,
        lsttxt    = \RSlsttxt
}

\newref{def}{
        name      = \RSdefntxt,
        names     = \RSdefnstxt,
        Name      = \RSDefntxt,
        Names     = \RSDefnstxt,
        rngtxt    = \RSrngtxt,
        lsttwotxt = \RSlsttwotxt,
        lsttxt    = \RSlsttxt
}

\newref{fact}{
        name      = \RSfacttxt,
        names     = \RSfactstxt,
        Name      = \RSFacttxt,
        Names     = \RSFactstxt,
        rngtxt    = \RSrngtxt,
        lsttwotxt = \RSlsttwotxt,
        lsttxt    = \RSlsttxt
}

\newref{nota}{
        name      = \RSnotationtxt,
        names     = \RSnotationstxt,
        Name      = \RSNotationtxt,
        Names     = \RSNotationstxt,
        rngtxt    = \RSrngtxt,
        lsttwotxt = \RSlsttwotxt,
        lsttxt    = \RSlsttxt
}

\newref{exa}{
        name      = \RSexampletxt,
        names     = \RSexamplestxt,
        Name      = \RSExampletxt,
        Names     = \RSExamplestxt,
        rngtxt    = \RSrngtxt,
        lsttwotxt = \RSlsttwotxt,
        lsttxt    = \RSlsttxt
}

\def\RSdefntxt{definition~}
\def\RSdefnstxt{definitions~}
\def\RSDefntxt{Definition~}
\def\RSDefnstxt{Definitions~}
\def\RSexampletxt{example~}
\def\RSexamplestxt{examples~}
\def\RSExampletxt{Example~}
\def\RSExamplestxt{Examples~}
\def\RSfacttxt{fact~}
\def\RSfactstxt{facts~}
\def\RSFacttxt{Fact~}
\def\RSFactstxt{Facts~}
\def\RSnotationtxt{notation~}
\def\RSnotationstxt{notations~}
\def\RSNotationtxt{Notation~}
\def\RSNotationstxt{Notations~}
\def\RSthmtxt{theorem~}
\def\RSthmstxt{theorems~}
\def\RSThmtxt{Theorem~}
\def\RSThmstxt{Theorems~}
\providecommand{\definitionname}{Definition}
\providecommand{\examplename}{Example}
\providecommand{\factname}{Fact}
\providecommand{\notationname}{Notation}
\providecommand{\theoremname}{Theorem}

\begin{document}
\global\long\def\norm#1{\left\Vert #1\right\Vert }%

\global\long\def\card#1{\left|#1\right|}%

\global\long\def\abs#1{\left|#1\right|}%

\global\long\def\nnz{\text{ nnz}}%

\global\long\def\nrows{\text{{\#rows}}}%

\global\long\def\ncols{\text{{\#cols}}}%

\global\long\def\flops{\mathcal{F}}%

\global\long\def\nflops{\text{{F}}}%

\global\long\def\nprocs{\text{{\ensuremath{n_{proc}}}}}%

\global\long\def\niters{\text{{\#iterations}}}%

\global\long\def\rank{\text{{rank}}}%

\global\long\def\supp{\text{{supp}}}%

\global\long\def\unif{\text{{Unif}}}%

\global\long\def\oi{\text{{OI}}}%

\global\long\def\io{\text{{Q}}}%

\global\long\def\st{\text{{\text{ s.t. }}}}%

\global\long\def\ov{\text{{\ensuremath{\vec{1}}}}}%

\global\long\def\sv{\text{{\ensuremath{\boldsymbol{\vec{s}}}}}}%

\global\long\def\pluseq{\text{{\ensuremath{\mathrel{+}}=}}}%

\title{Information-Theoretic Communication Lower Bounds for Sparse Matrix
Multiplication}
\author{Grace Dinh}
\authornote{Work done in part while the author was a research fellow at
the Simons Institute for the Theory of Computing, UC Berkeley.}
\email{gndinh@cornell.edu}
\affiliation{\institution{Cornell University} \city{Ithaca} \state{New York}
\country{USA} }
\author{Vivienne Goyal}
\email{vivienne.goyal@princeton.edu}
\affiliation{\institution{Princeton University} \city{Princeton} \state{New Jersey}
\country{USA} }
\author{Raye Kimmerer}
\email{kimmerer@princeton.edu}
\affiliation{\institution{Princeton University} \city{Princeton} \state{New Jersey}
\country{USA} }
\author{James Demmel}
\affiliation{\institution{UC Berkeley} \city{Berkeley} \state{California} \country{USA} }
\email{demmel@berkeley.edu}
\author{Zhiru Zhang}
\email{zhiruz@cornell.edu}
\affiliation{\institution{Cornell University} \city{Ithaca} \state{New York}
\country{USA} }
\begin{abstract}
The communication cost of sparse-sparse matrix multiplication (SpGEMM)
is heavily influenced by the sparsity structure of its input matrices,
even when their dimensions and nonzero counts are fixed. We present
an information-theoretic method for deriving \emph{instance-specific}
communication lower bounds in both sequential and parallel models.
Our approach incorporates a variety of sparsity statistics as entropic
inequalities in an optimization program that can be solved either
numerically or symbolically; the more information about the sparsity
structure is supplied, the stronger the bounds.

We also compare our bounds to measured cache traffic of several
state-of-the-art libraries and find that our bounds are significantly
closer to the measured traffic than previous bounds, typically by
more than a factor of two; they also predict measured traffic significantly better than both previous bounds and statistical models that assume Erdős--Rényi structure.
\end{abstract}
\maketitle

\section{Introduction}

Sparse matrix-matrix multiplication (SpGEMM) forms an important computational
bottleneck in applications as varied as graph algorithms, algebraic
multigrid, simulation \citet{BBD+13-Communication,KG11-Graph},
machine learning \citet{ZGD+21-Big,LJZT24-Sparser}, and computational
biology \citet{Gui22-Parallel}. This has motivated substantial work
on efficient algorithms, libraries, and hardware for sparse matrix
products \citet{GJC+23-Systematic,SJL+20-MatRaptor,NOU+23-TeAAL}.

In most cases, the cost of SpGEMM is dominated by data movement -
either between levels of the memory hierarchy, or between processors
in a parallel setting - which can significantly outweigh the cost
of arithmetic \citet{BDKS16-Hypergraph,KK20-GAMMA}. For such communication
bound problems, reducing data movement \emph{and determining to what
extent data movement can be reduced} are central to optimizing and
evaluating software implementations, as well as to hardware design-space
exploration, where communication lower bounds can guide high-level
architectural design decisions \citet{ESR+15-Using,HTEP24-Mind}.

Unlike with \emph{dense} linear algebra, whose regular structure allows
for straightforward communication bounds based on matrix dimensions
and hardware parameters \citet{HK81-complexity,ITT04-Communication,KKB+19-Redblue,SLLv19-Tight},
SpGEMM's communication costs can vary widely depending on the exact
location of its nonzeros, which we refer to as \emph{sparsity structure}\footnote{some papers, e.g. \citet{Gre12-Sparse}, use the term \emph{conformation}.};
even products with identical dimensions and nonzero counts can require
different amounts of communication \citet{BDKS16-Hypergraph}. 

Previous work on finding sparsity-sensitive communication lower bounds
for both serial and parallel SpGEMM has had several major impediments
to their practical application. These include worst-case bounds over
\emph{all} matrices that share the same number of inputs and outputs
\citet{PS14-Input} (which therefore do not necessarily apply to a
matrix of interest, and can in fact be broken by real-world implementations,
as we will see in Section \ref{sec:Background-and-Motivation}); non-universal
bounds over specific families of algorithms \citet{Gre12-Sparse};
bounds only applicable to uniform Erdős--Rényi matrices \citet{BBD+13-Communication};
and bounds parametrized by computationally intractable structure-specific
partitioning computations \citet{BDKS16-Hypergraph}.

This paper presents, to our knowledge, the first \emph{universal},
\emph{practically computable, structure-dependent} communication lower
bounds for SpGEMM (and similarly structured tensor contractions).
Specifically, our contributions are as follows:
\begin{itemize}
\item We develop a novel method, building on techniques from information
theory and database cardinality estimation, of \emph{programmatically}
finding communication lower bounds for nested loops accessing arrays.
These bounds can both be symbolically derived in terms of statistics
of matrix structure and numerically computed efficiently.
\item Using this method, as well as some new information theoretic inequalities
parametrized by matrix sparsity structure, we derive new communication
lower bounds for SpGEMM in both sequential and parallel machine models.
These bounds can (but do not require) incorporate additional information
about the skewness and layout of the sparsity structures of the matrices,
which are described using compact, intuitive matrix statistics.
\item We compare these lower bounds to both previous bounds and to state-of-the-art
implementations of SpGEMM to show that they can effectively serve
to bound and estimate performance \emph{in practice}.
\end{itemize}

\section{Background and Motivation\label{sec:Background-and-Motivation}}

\subsection{Problem and machine model}

Let $A$ and $B$ be matrices of dimensions $L_{i}\times L_{j}$ and
$L_{j}\times L_{k}$, respectively, over a commutative semiring. Let
$S_{A}$ be the set of indices $(i,j)$ such that $A[i,j]\ne0$ (analogously
for $S_{B}$), so $\nnz A=\left|S_{A}\right|$. We will consider the
SpGEMM operation $C=A\times B$:
\begin{align}
 & \text{for }(i,j,k)\in\mathcal{\flops}:\nonumber \\
 & \qquad C[i,k]\pluseq A[i,j]\times B[j,k].\label{eq:matmulloops}
\end{align}
where $\mathcal{F}$ is the set of \emph{nontrivial flops} $\flops\coloneqq\{(i,j,k):(i,j)\in S_{A},\ (j,k)\in S_{B}\}$.

Following standard practice for sparse matrix multiplication \citet{PS14-Input,GJ10-Complexity},
we will permit the execution of the loop iterations of (\ref{eq:matmulloops})
in any order, or \emph{schedule}, but we will exclude algorithms that
perform unnecessary recomputation as well as Strassen-like algorithms
that rely on numerical cancellation, which are both generally unsuited
for sparse matrix multiplications (as they introduce unnecessary fill-in)
and are incompatible with arbitrary semirings \citet{PS14-Input}.
As a result, for a given $A$ and $B$, the number of flops (indexed
by $\left(i,j,k\right)$) is fixed to the number of loop iterations $\nflops=\left|\flops\right|$
(we will consider a fused multiply-add as a single flop, and therefore
treat loop iterations and flops interchangeably).

We will use the following shared-memory external memory model: we
have $n_{proc}$ processors, each of which is connected to its own
\emph{fast memory} capable of storing $M$ matrix elements, or \emph{words}.
Processor $p$ can only read from and write to locations on its own
fast memory. All fast memories are connected to a single shared \emph{slow
memory}, which holds the input at the beginning of execution. 
Matrix elements or partial sums can be transferred between fast and
slow memory; each individual value transferred incurs a communication
cost of $1$. The \emph{total communication cost} $Q$ is the total
amount of data moved between all fast memories and the slow memory;
and the \emph{per-processor communication }cost $Q_{p}$ denotes the
amount of data moved between processor $p$ and slow memory.

When $\nprocs=1$, this becomes the sequential two-level I/O model
of \citet{HK81-complexity}. Communication lower bounds in this model
also imply communication bounds in a distributed-memory model, as
any distributed-memory execution can be simulated on shared memory
by turning a processor-to-processor message into a write by the sender
and a read by the recipient.

\subsection{Prior Work}

Several prior studies have attempted to bound the communication cost of SpGEMM.
Bounds ported directly from dense linear algebra, such as $\nflops/\sqrt{M}$
\citep{BDHS11-Minimizing}, are extremely loose in practice.

\citet{BDKS16-Hypergraph} provides instance-specific communication
lower bounds in both the serial and distributed parallel settings
(but not to the shared-memory parallel model, which is explicitly
left as future work). However, these bounds are expressed as a minimization
problem over \emph{all} possible partitionings of a hypergraph induced
by the sparsity structures of the input and output matrices - either
into $\nprocs$ partitions in the parallel case, or into partitions
of bounded memory footprints for the serial case. Unfortunately, hypergraph
partitioning in general is NP-hard, and no tractable simplification
is known for the special case of SpGEMM hypergraphs (the NP-hardness
of this special case is left open), and heuristic partitioners do
\emph{not} generate communication lower bounds; as a result, these
bounds are impractical to compute. Even if this were to be overcome,
because the bounds depend on the \emph{complete} sparsity pattern
of the input matrices, they are difficult to generalize to statements
about \emph{families} of matrices: a bound for one instance does not
directly imply a bound for another with similar structure.

Other works either operate in a different model entirely or make assumptions
about the algorithm or sparsity structure of the input and output
matrices.

In the \emph{sequential} case, \citet{PS14-Input} shows that given
some integers $N$ and $Z$, there exist some matrices $C=AB$ such
that $\nnz A,\nnz B\le N$, $\nnz C\le Z$, and any algorithm computing
$C=A\times B$ requires at least
\[
\Omega\left(N\min\!\left\{ \frac{N}{M},\sqrt{\frac{Z}{M}}\right\} \right)
\]
communication. This is not an instance-specific bound, but rather
a \emph{worst-case} bound over \emph{all} possible matrices with the
same number of nonzeros; in fact, our experiments in Section \ref{sec:empirical}
show that these bounds are sufficiently pessimistic that they predict
costs far \emph{more} expensive than what standard implementations
can achieve in practice.

In the parallel shared-memory case, \citet{Gre12-Sparse} derives
similar \emph{worst-case} bounds for SpGEMM, but such bounds are only \emph{guaranteed}
to hold for so-called ``pseudo-rectangular'' algorithms
that update one $M$-bounded rectangular submatrix of the output at
a time; the bound's general applicability is conjectured but (to our
knowledge) remains open. While our experiments do not show these bounds
to be as pessimistic as the serial bounds of \citet{PS14-Input},
they are often significantly weaker than our bounds in practice.

Alternatively, some other previous work, e.g. \citet{BBD+13-Communication},
makes statistical assumptions about the sparsity structure of the
matrices - most commonly uniform/Erdős--Rényi sparsity, uncorrelated
between inputs. While such statistical assumptions can lead to useful
performance models, e.g. \citet{WTP+22-Sparseloop,SXZ+25-Probabilistic},
such assumptions cannot be assumed to hold for matrices with structure
and can lead to overly optimistic or pessimistic figures, as we will
demonstrate in Section \ref{sec:empirical}.

\section{I/O Complexity Bounds through Information Theory}

\subsection{Preliminaries}

To motivate our approach, let us first use information theory to rederive
the classical \emph{serial} $\Omega\left(\nflops/\sqrt{M}\right)$
communication lower bound of \citet{HK81-complexity}.

Any schedule that executes (\ref{eq:matmulloops}) must execute all
$\nflops$ of its loop iterations. Following the approach of \citet{CDK+13-Communication},
partition the schedule into sequences of instructions (i.e., subsets
of $\flops$) during whose execution $M$ words of data are moved
between fast and slow memory; we refer to each such sequence as a
\emph{segment}. Let $\min\left|s\right|$ denote the minimum number
of $M$-segments required to partition (\ref{eq:matmulloops}). Then
the communication $Q$ required to execute (\ref{eq:matmulloops}) is bounded
below by 
\begin{equation}
\io\ge M\left(\min\left|s\right|-1\right)\label{eq:io}
\end{equation}
since every segment after the first requires us to move $M$ words;
the problem therefore reduces to finding a lower bound on $\min\left|s\right|$.

We note that at the beginning of each segment, fast memory contains
$M$ words; during the execution of the segment, an additional $M$
words may be accessed. As a result, the number of words that may be
accessed during a segment is at most $2M$. One way of determining
such a lower bound is to find an \emph{upper} bound on the number
of flops that can be executed in a single segment, which gives 
\begin{equation}
\min\left|s\right|\ge\frac{\nflops}{\text{max flops/segment}}\ .\label{eq:seg-upper-bound}
\end{equation}
The standard approach in this is to use geometric inequalities, such
as that of Loomis and Whitney \citet{LW49-inequality}, to bound the
maximum number of flops in a segment to $O\left(M^{3/2}\right)$. 

Alternatively, one can view this problem as one of finding a worst-case
cardinality bound for a database join. Let $R_{A}$, $R_{B}$, and
$R_{C}$ be sets, or \emph{relations}, consisting of the addresses
of the entries of $A$, $B$, and $C$ used by a segment; these relations
contain $ij$, $jk$, and $ik$ index pairs, respectively. The flops
(i.e. tuples $\left(i,j,k\right)$) executed by the segment form a
subset of the triangle join $R_{A}\Join R_{B}\Join R_{C}$;
as a result, it suffices to find an upper bound on the cardinality
of the triangle join subject to constraints on the cardinalities of
its component relations (by construction, each of these relations
has cardinality no more than $2M$). This bound can be manually
derived, as in \citet{AGM13-Size}, or programmatically discovered
\citet{ANS17-What,CBS19-Pessimistic} through an optimization problem.
As the remainder of this paper will demonstrate, the latter approach
can be much more easily extended to apply to parallel models or to
incorporate information about sparsity. We review it in the remainder
of the subsection.

We first establish some preliminaries:
\begin{defn}[Shannon entropy]
\label{def:entropy-def}Let $D$ be a joint distribution of discrete
random variables $\mathscr{X}=\{x_{1},\ldots,x_{n}\}$, each taking values in $\mathbb{N}$.
For some subsets of random variables $\mathscr{I},\mathscr{J}\subseteq\mathscr{X}$,
let $D_{\mathscr{I}}$ be the marginal distribution of $D$ on $\mathscr{I}$,
and define its Shannon entropy by
\[
h_{D}(\mathscr{I})\coloneqq-\sum_{a\in\supp(D_{\mathscr{I}})}D_{\mathscr{I}}(a)\lg D_{\mathscr{I}}(a),
\]
where $\lg$ denotes the base-2 logarithm and $D_{\mathscr{I}}(a)$
denotes the probability that the random variables in $\mathscr{I}$ take
on value $a$ under $D_{\mathscr{I}}$. The entropy of $\mathscr{I}$
conditioned on $\mathscr{J}$ is defined as 
\[
h_{D}(\mathscr{I}\mid\mathscr{J})\coloneqq h_{D}(\mathscr{I}\cup\mathscr{J})-h_{D}(\mathscr{J})\ .
\]

When it is clear what the underlying distribution is, we will omit
the subscript and simply write $h(\mathscr{I})$. Following the convention
of previous work \citet{ANS17-What,CBS19-Pessimistic,ANOS24-Join},
we will omit the commas and brackets to denote the entropy for a set
of random variables; e.g. we will write $h(xy)$ instead of $h(\{x,y\})$.
\end{defn}

A straightforward convexity argument gives the following useful property
\citet{Yeu02-First}:

\begin{fact}
\label{fact:supportbound}For any distribution $D$ and subset of
its random variables $\mathscr{I}$, $h_{D}(\mathscr{I})\le\lg|\supp(D_{\mathscr{I}})|$.
The inequality is tight if and only if the marginal is uniform.
\end{fact}

We can use this to provide the desired bound on the maximum number
of flops achievable in a segment. To do so, fix an arbitrary nonempty
segment and consider the uniform distribution $D^{seg}$ over tuples
$(i,j,k)$ in the segment, each representing a flop. Our goal will
be to upper bound the size of the support of $D^{seg}$, or, equivalently
(following \Factref{supportbound}), $h_{D^{seg}}(ijk)=\lg\left|\supp\left(D^{seg}\right)\right|$.

If flop $(i,j,k)$ is in the segment, $A[i,j]$ must be accessed during
the segment. As a result, at least $\card{\supp(D^{seg}_{\{i,j\}})}$ distinct
elements of $A$ are accessed in the segment; by construction, this
number is bounded by $2M$. Applying \Factref{supportbound} gives
\begin{equation}
h_{D^{seg}}(ij)\le\lg\card{\supp(D^{seg}_{\{i,j\}})}\le\lg(2M).\label{eq:a-constr}
\end{equation}

Applying similar reasoning to $B$ and $C$ gives 
\begin{eqnarray}
h_{D^{seg}}(jk) & \le & \lg\left(2M\right)\label{eq:bc-constr}\\
h_{D^{seg}}(ik) & \le & \lg\left(2M\right)\nonumber 
\end{eqnarray}

Using (\ref{eq:a-constr}) and (\ref{eq:bc-constr}) to bound $h_{D^{seg}}(ijk)$
will require the use of the Shannon inequalities, whose proof can
be found in any reference on information theory, e.g. \citet{Yeu02-First}.
\begin{fact}[Shannon inequalities]
\label{thm:shannon-ineqs}For any distribution $D$ and sets of random
variables $\mathscr{I},\mathscr{J}\subseteq\mathscr{X}$ as defined
in \Defref{entropy-def}, the following statements hold
\begin{itemize}
\item $h_{D}\left(\emptyset\right)=0$ 
\item Nonnegativity: $h_{D}(\mathscr{I}\vert\mathscr{J})\ge0$ 
\item Submodularity: $h_{D}\left(\mathscr{I}\cup\mathscr{J}\right)+h_{D}\left(\mathscr{I}\cap\mathscr{J}\right)\le h_{D}\left(\mathscr{I}\right)+h_{D}\left(\mathscr{J}\right)$ 
\end{itemize}
\end{fact}

These inequalities can be exhaustively enumerated by assigning every
element of the power set of $\left\{ i,j,k\right\} $ to $\mathscr{I}$,
$\mathscr{J}$. Combining the enumerated Shannon inequalities with
(\ref{eq:a-constr}) and (\ref{eq:bc-constr}) gives us a linear set
of constraints over all possible joint entropies over $i$, $j$,
$k$, parameterized by $\lg\left(2M\right)$. Maximizing $h_{D^{seg}}\left(i,j,k\right)$
over this set of constraints is a \emph{parametric} linear program
(LP), which can be solved numerically using a standard LP solver for
a given value of $M$. It is also possible to symbolically solve for
$\max h_{D^{seg}}\left(i,j,k\right)$ as a piecewise linear function
of the parameter $\lg\left(2M\right)$ using multiparametric linear
program (MPLP) algorithms, e.g. \citet{BBM03-Geometric}. Using the
MPLP solver in the MPT3 library \citet{HKJM13-MultiParametric} gives
a single piece solution: $\max h_{D^{seg}}\left(i,j,k\right)=\frac{3}{2}\lg\left(2M\right)$.
As $D^{seg}$ is uniform, \Factref{supportbound} immediately implies
that $\card{\supp\left(D^{seg}\right)}\le\left(2M\right)^{3/2}$, and substituting
into (\ref{eq:io}) gives the desired $\Omega\left(\nflops/\sqrt{M}\right)$
bound.

This bound is asymptotically tight for dense matrix multiplication,
and can be achieved by tiling the loop nest (\ref{eq:matmulloops})
with square tiles of dimension $\Theta\left(\sqrt{M}\right)$, but
is known to be extremely loose for sparse matrix multiplication \citet{BDKS16-Hypergraph}.

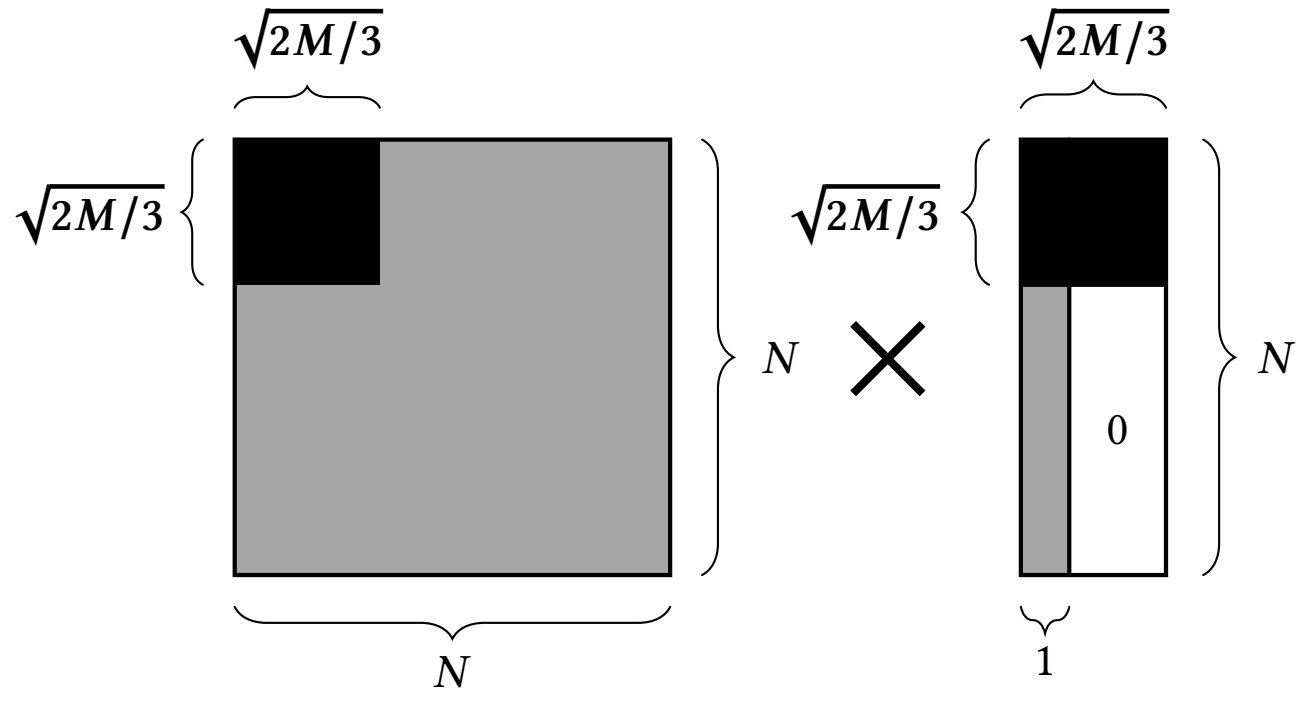
\begin{figure}
\begin{tikzpicture}[x=1cm,y=1cm, line cap=round,scale=0.8]

\def\W{1.2}    
\def\H{3.6}    
\def\wone{0.4}
\def\htop{1.2}
\def\N{3.6}
\def\dist{2.9}
\def\Msq{1.2}

\begin{scope}[shift={({-(\N+\dist)},0)}]
    \fill[black!35] (0,0) rectangle (\N,\N);
    
    \fill[black] (0,\N-\Msq) rectangle (\Msq,\N);
    
    \draw[thick] (0,0) rectangle (\N,\N);

    
    \draw[decorate,decoration={brace,amplitude=6pt,raise=6pt,mirror}]
      (0,0) -- (\N,0)
      node[midway,below=12pt] {$N$};

    \draw[decorate,decoration={brace,amplitude=6pt,raise=6pt, mirror}]
      (\N,0) -- (\N,\N)
      node[midway,right=14pt] {$N$};

    \draw[decorate,decoration={brace,amplitude=4pt,raise=6pt}]
      (0,\N) -- (\Msq,\N)
      node[midway,above=10pt] {$\boldsymbol{\sqrt{2M/3}}$};
      
    \draw[decorate,decoration={brace,amplitude=4pt,raise=6pt}]
      (0,\N-\Msq) -- (0,\N)
      node[midway,left=10pt] {$\boldsymbol{\sqrt{2M/3}}$};
\end{scope}

\node[scale=3] at ({-\dist/2+ 0.35}, {\H/2}) {$\times$};

\fill[black!35] (0,0) rectangle (\wone,\H-\htop); 
\fill[black] (0,\H-\htop) rectangle (\W,\H); 

\draw[thick] (0,0) rectangle (\W,\H);
\draw[thick] (\wone,0) -- (\wone,\H);          
\draw[thick] (0,\H-\htop) -- (\W,\H-\htop);    

\node at ({(\wone+\W)/2},{(\H-\htop)/2}) {$0$};

\draw[decorate,decoration={brace,amplitude=5pt,raise=6pt}]
  (0,\H) -- (\W,\H)
  node[midway,above=10pt] {$\boldsymbol{\sqrt{2M/3}}$};

\draw[decorate,decoration={brace,amplitude=5pt,raise=6pt}]
  (0,\H-\htop) -- (0,\H)
  node[midway,left=12pt] {$\boldsymbol{\sqrt{2M/3}}$};

\draw[decorate,decoration={brace,amplitude=6pt,raise=7pt,mirror}]
  (\W,0) -- (\W,\H)
  node[midway,right=14pt] {$N$};

\draw[decorate,decoration={brace,amplitude=5pt,raise=6pt,mirror}]
  (0,0) -- (\wone,0)
  node[midway,below=10pt] {$1$};

\end{tikzpicture}\caption{\label{fig:pathological-segment-size}
A dense-sparse matrix multiplication
whose I/O complexity is poorly captured by the classical segment-size
based bound. Both the black and the grey shaded areas represent nonzeros.
The black shaded area corresponds to the data accessed by a $M^{3/2}$
size segment. The white area represents zeros.}
\end{figure}

In fact, \emph{any} communication lower bound based on bounding segment
\emph{size} is insufficient for sparse multiplications, as the shape
of the problem may be nonuniform: one part of the computation may
allow for large segments that cannot be used for other parts of the
computation. Consider, for example, the problem of multiplying a dense
$N\times N$ matrix $A$ by an $N\times\sqrt{2M/3}$ sparse matrix
$B$ whose sparsity structure is described in Figure \ref{fig:pathological-segment-size},
for $N\gg\sqrt{M}$. The maximum segment size remains $O\left(M^{3/2}\right)$,
attained by multiplying a $\sqrt{2M/3}\times\sqrt{2M/3}$ submatrix
of $A$ with the dense $\sqrt{2M/3}\times\sqrt{2M/3}$ submatrix of
$B$; therefore applying (\ref{eq:seg-upper-bound}) gives the standard
I/O lower bound of $\Omega\left(\nflops/\sqrt{M}\right)$. This is,
however, an extremely weak bound: as the rest of the problem outside
the (inconsequentially small, given our assumption that $N\gg\sqrt{M}$)
dense segment is a sparse matrix-\emph{vector} multiplication, with
no data reuse; hence the I/O lower bound is in fact $\Omega\left(\nflops\right)$.

As a result, in order to accommodate irregularly structured sparse
computations, we must change the underlying optimization program to
consider the entire iteration space \emph{globally} instead of looking
at one local segment at a time.

\subsection{Serial Bounds}

Instead of trying to bound the size of \emph{a single} segment, we
will instead directly bound the minimum number of segments in any
partitioning of the entire iteration space. Specifically, given an
arbitrary partitioning of the loop nest into segments, we assign each
loop iteration $(i,j,k)$ a \emph{segment ID} $s\in\mathbb{N}$. Two
loop iterations are in the same segment if and only if their segment
IDs are the same. Our goal will be to find a lower bound for the number
of distinct values of $s$.

To do so, consider $s$ as a random variable alongside $i$, $j$,
and $k$. Our general strategy will be to find a lower bound on the
entropy of $s$ under some distribution, which will (by \Factref{supportbound})
provide a lower bound on the number of unique values of $s$.

Our underlying distribution $D^{serial}$ will be the uniform distribution
over all possible values of $\left(i,j,k,s\right)$ such that $\left(i,j,k\right)$
is a loop iteration and $s$ is the segment ID corresponding to that
loop iteration. For the remainder of this subsection, all entropies
(unless otherwise stated) will be taken with respect to this distribution,
and we will therefore omit the subscript on the entropy values $h$.
Notice that the marginalization of this distribution to $\left(i,j,k\right)$
is also uniform, as each loop iteration $\left(i,j,k\right)$ is assigned
exactly one segment ID. Therefore:
\begin{align}
h\left(ijk\right) & =\lg\nflops\ .\label{eq:hijk-flops}
\end{align}
Furthermore, as each loop iteration is assigned exactly one $s$-value,
we have 
\begin{equation}
h\left(s\vert ijk\right)=0\ .\label{eq:s|ijk}
\end{equation}

As our distribution is now global over \emph{all} flops, rather than
over the flops in a single segment, the inequalities (\ref{eq:bc-constr})
must be modified to \emph{condition} on the segment ID $s$, which
leads to the constraints:
\begin{eqnarray}
h\left(ij\vert s\right) & \le & \lg\left(2M\right)\nonumber \\
h\left(jk\vert s\right) & \le & \lg\left(2M\right)\label{eq:conditional-mem-constraints}\\
h\left(ik\vert s\right) & \le & \lg\left(2M\right)\nonumber 
\end{eqnarray}

Our main serial bound is given by the following:
\begin{thm}
\label{thm:serial-bound}Let $\tilde{h}(s)$ be the minimum value
of $h(s)$ subject to the constraints (\ref{eq:hijk-flops}), (\ref{eq:s|ijk}),
and (\ref{eq:conditional-mem-constraints}). Then $M\left(2^{\tilde{h}(s)}-1\right)$
is a lower bound for the communication required to perform the matrix
multiplication (\ref{eq:matmulloops}). We may additionally strengthen
the bound by imposing additional entropic inequalities solely implied
by the sparsity structure of $A$, $B$, or $C$ (i.e. not dependent
on the schedule).
\end{thm}

\begin{proof}
For any valid schedule with $t$ segments, the entropies of the induced
distribution $D^{serial}$ satisfy all the stated constraints, so
$\tilde{h}(s)\le h(s)\le\lg t$. The desired communication lower bound
$\io\ge M(t-1)\ge M\left(2^{\tilde{h}(s)}-1\right)$ follows immediately
from (\ref{eq:io}).

As the additional inequalities depend only on the sparsity structure
of $A$, $B$, and $C$, they hold for every valid schedule; as a
result, imposing them preserves feasibility of the induced distributions
and cannot decrease the minimum of $h(s)$. We will provide examples
of these inequalities - and the terms they induce in the communication
lower bound - in Section \ref{sec:Sparse-Bounds}.
\end{proof}

We note that the constant factors in this bound are not tight. Following
the approach of \citet{SLLv19-Tight}, they can be tightened by redefining
a segment to correspond to $R$ words of data movement, reworking
(\ref{eq:conditional-mem-constraints}) so that the constraint reads
$2^{h\left(ij\vert s\right)}+2^{h\left(jk\vert s\right)}+2^{h\left(ik\vert s\right)}\le M+R$,
solving the resulting nonlinear optimization problem, and then optimizing
over $R$. As the resulting constraint is no longer linear, symbolically
solving the resulting optimization problem (especially once augmented
with the sparsity-related constraints introduced in the next section)
is a significantly more complex and slow process; we leave implementing
this to future work and focus on the big-$\Omega$ estimates for this
paper.

We can, however, \emph{numerically} optimize this quantity, as nonnegativity of entropy means that the optimization problem
\emph{does} remain convex (and thus numerically solvable). Several sparsity constraints that will be introduced in Section \ref{sec:Sparse-Bounds} will turn out to be redundant in the relaxed memory constraint setting with individual $2M$-bounds on the memory footprints of each matrix, but do improve the numeric bounds under the constraint
\begin{equation}
2^{h\left(ij\vert s\right)}+2^{h\left(jk\vert s\right)}+2^{h\left(ik\vert s\right)}\le 2M.\label{eq:summem}
\end{equation}

\subsection{Parallel Bounds}

We can apply similar techniques to find both total and per-processor
bounds on communication in the parallel setting. To do so, individually
partition the set of instructions performed by each processor into
segments where $M$ words are communicated to shared memory, and give
all segments globally distinct IDs. Then each segment $s$ is assigned
to a single processor $p\in[1,\nprocs]$, which we treat as an additional
random variable. This maps each parallel execution schedule to a set
of $(i,j,k,s,p)$ tuples, indicating that loop iteration $\left(i,j,k\right)$
is executed in segment $s$ on processor $p$. Let $D^{par}$ be the
uniform distribution on this set of tuples; all entropies below refer
to this distribution.

The serial constraints (\ref{eq:hijk-flops}), (\ref{eq:s|ijk}),
and (\ref{eq:conditional-mem-constraints}) continue to hold, but
we must add the following constraints to ensure that each segment
is assigned to only one processor, and that the number of distinct
values of $p$ is constrained by the number of processors on the system:
\begin{eqnarray}
h(p) & \le & \lg\nprocs\label{eq:parallel-constraints}\\
h(p\mid s) & = & 0\nonumber 
\end{eqnarray}
\begin{thm}
\label{thm:parallel-bound}Let $\tilde{h}(s|p)$ be the minimum value
of $h(s|p)$ subject to (\ref{eq:hijk-flops}), (\ref{eq:s|ijk}),
(\ref{eq:conditional-mem-constraints}), and (\ref{eq:parallel-constraints}).
Then for every valid parallel execution that runs at least one flop
on each processor, the maximum per-processor communication satisfies
\begin{align}
\io_{\max}\coloneqq\max_{p}\io_{p} & \ge M\left(2^{\tilde{h}(s\mid p)}-1\right).\label{eq:parallel-total-bound}
\end{align}
Additional entropic inequalities implied by the sparsity structure
and valid for every schedule may be imposed in the minimization.
\end{thm}

\begin{proof}
For any valid parallel schedule, the conditional entropy $h\left(s\vert p\right)$
is the weighted average
\[
h(s\mid p)=\sum_{p^{\dagger}}\Pr\left[p=p^{\dagger}\right]h(s\mid p=p^{\dagger})
\]
so there must exist some processor $p'$ for which
\[
h(s\mid p=p')\ge h(s\mid p)\ge\tilde{h}(s\mid p)\ .
\]
As the number of segments on $p'$ is bounded below by $2^{h(s\mid p=p')}$,
the desired communication lower bound follows.
\end{proof}

For dense multiplication, this recovers the familiar memory-dependent
lower bound 
\begin{equation}
\Omega\left(\nflops/(\nprocs\sqrt{M})\right)\label{eq:basic-parallel-bound}
\end{equation}
 for $\io_{\max}$ when $\nflops\gg\nprocs M^{3/2}$ \citet{SD11-CommunicationOptimal}.

The marginal distribution on $(i,j,k)$ is unchanged from the serial
construction, so constraints involving only these variables apply
to both minimizations. Section \ref{sec:Sparse-Bounds} introduces
such inequalities from the sparsity structure of $A$, $B$, and $C$,
together with the resulting communication bounds.

\section{Sparse Bounds\label{sec:Sparse-Bounds}}

In order to find nontrivial \emph{sparse} lower bounds, we must incorporate
information about the sparsity structure of $A$, $B$, and $C$.
We do so by adding additional entropic constraints parametrized by
the matrices' sparsity structure. Any subset of these constraints
(based on what sparsity statistics are available) can be straightforwardly
plugged into the optimization problems of Theorems \ref{thm:serial-bound}
and \ref{thm:parallel-bound} to provide stronger serial and parallel
lower bounds.

This section presents several such constraints and the lower bounds
that can be derived from them, and analyzes the impact of these constraints
with real-world matrices. We expect that additional entropic inequalities
will be discovered to describe specific sparse structures, leading
to stronger lower bounds.

\subsection{Dimension and nnz constraints}

The most straightforward inequalities are those based on the matrix
dimensions.

As $A$ is $L_{i}\times L_{j}$ and $B$ is $L_{j}\times L_{k}$,
$i$ can take only $L_{i}$ distinct values (and analogously for $j$
and $k$). Therefore:
\begin{eqnarray}
h(i) & \le & \lg L_{i}\nonumber \\
h(j) & \le & \lg L_{j}\label{eq:loop-bounds}\\
h(k) & \le & \lg L_{k}\nonumber 
\end{eqnarray}
If $A$ has rows that are entirely zero, or rows that contribute nothing
to the output, $L_{i}$ can be replaced with the number of rows with
at least one nonzero that contributes to the product.

Introducing these constraints gives us a serial communication lower
bound of $\Omega\left(F/\min\left(L_{i},L_{j},L_{k}\right)\right)$
(on top of the $F/\sqrt{2M}$ bound); in the dense case, this matches
the size of the input.

Furthermore, the set of $\left(i,j\right)$-tuples such that there
exists some $k$ such that $\left(i,j,k\right)$ is a nontrivial flop
is, by definition, a subset of the nonzero elements of $A$, which
implies 
\begin{equation}
h\left(i,j\right)\le\lg\nnz A\ .\label{eq:nnz-ij-constr}
\end{equation}
Similar logic implies that $h\left(j,k\right)\le\lg\nnz B$ and $h\left(i,k\right)\le\lg\nnz C$.
This leads to the following additional communication lower bounds:
\[
\frac{F}{\sqrt{\rho}\nprocs},\frac{F^{2}\rho}{\nprocs^{2}\nnz A\nnz B\nnz C}
\]
where $\rho=\min\left(\nnz A,\nnz B,\nnz C\right)$.

\subsection{Nonzeros per row and column}

We can also take advantage of more specific information about the
sparsity structure of our matrices, which we encode through \emph{degree}
constraints. 
\begin{notation}
Let $x$ and $y$ be random variables with finite support. Then let
\emph{degree} or \emph{cardinality} of $x$, denoted $\deg\left(x\right)$,
be the number of distinct values that $x$ can take, i.e. $\left|\supp\left(x\right)\right|$.
The degree of $x$ with respect to $y$, $\deg\left(x\vert y\right)$,
is a vector indexed by all possible values of $y$ (i.e. of length
$\deg\left(y\right)$), each of whose elements is the number of unique
values of $x$ that may occur when $y$ is restricted to its index. 
\end{notation}

For example, $\deg\left(i\right)$ is the number of rows of $A$ that
contain at least one nonzero element, and $\deg(i\vert j)$ is a vector
describing how many nonzero elements are present in each of the $j$
columns of $A$.

Upper bounds on the number of nonzeros per row (or column) of a matrix
straightforwardly give bounds on the conditional entropy of the associated
indices. For example, one can bound the maximum entropy of $k$ given
$j$ with the log of the maximum number of nonzeros per row of $B$,
which can be written using the infinity norm of the degree vector:
\begin{eqnarray}
h\left(k\vert j\right) & \le & \lg\max\deg_{B}\left(k\vert j\right)\nonumber \\
 & = & \lg\left\Vert \deg_{B}\left(k\vert j\right)\right\Vert _{\infty}\label{eq:infty-norm}
\end{eqnarray}

Furthermore, if we know the maximum number of nonzeros per row and
column of $A$, $B$, or $C$ (let the maximum of these quantities
be denoted $z$), we obtain the additional bound 
\[
\Omega\left(\frac{\nflops}{z\nprocs}\right)\ .
\]

These can also lead to bounds in concert with those resulting from the nnz constraints, such as:
\begin{equation}
\frac{\nflops}{\norm{\deg_A(i\mid j)}_{\infty} \nprocs} \label{linf-bound-eg}
\end{equation}

and 
\begin{equation}
\frac{\nflops}{\sqrt{\nnz A} \nprocs} \label{nnz-bound-eg}
\end{equation}




However, maximum degree is a coarse metric. Many sparse
matrices, such as those representing social network graphs, have significant
variations in the number of nonzero elements per row or column; this
is poorly captured by both total-nnz bounds and max-degree bounds.
To capture the \emph{distribution} of nonzeros per row and/or column,
instead of only its maximum, we use the following theorem from \citet{ANOS24-Join},
which generalizes (\ref{eq:infty-norm}) from the infinity norm to
\emph{arbitrary $L_{\rho}$} norms of the cardinality vector:
\begin{thm}
\label{thm:lp-norm-bound}For any $\rho\ge1$ and random variables
$x$, $y$, 
\[
h\left(x\right)+\rho\cdot h\left(y\vert x\right)\le\lg\left\Vert \deg\left(y\vert x\right)\right\Vert ^{\rho}_{\rho}
\]
\end{thm}

In practice, we found that these norm constraints can sometimes, but not always, significantly strengthen the lower bounds obtained under the summed memory constraint (\ref{eq:summem}). In our experiments, we found that the degree of improvement in the
lower bound obtained by introducing $L_{2}$ and $L_{3}$ norm inequalities
is sensitive to the size of fast memory $M$ relative to the size
of the input matrices.

Specifically, writing $N=\nnz A+\nnz B$, we found a ``nnz-per-row/column skew sensitive''
regime (for our datasets, roughly $M\in[0.1N,0.7N]$) within which
introducing norm inequalities greatly strengthened our bounds. For
significantly larger or smaller fast memory sizes, adding norm inequalities
provided almost no improvements.

\begin{figure}
\includegraphics[width=1\columnwidth]{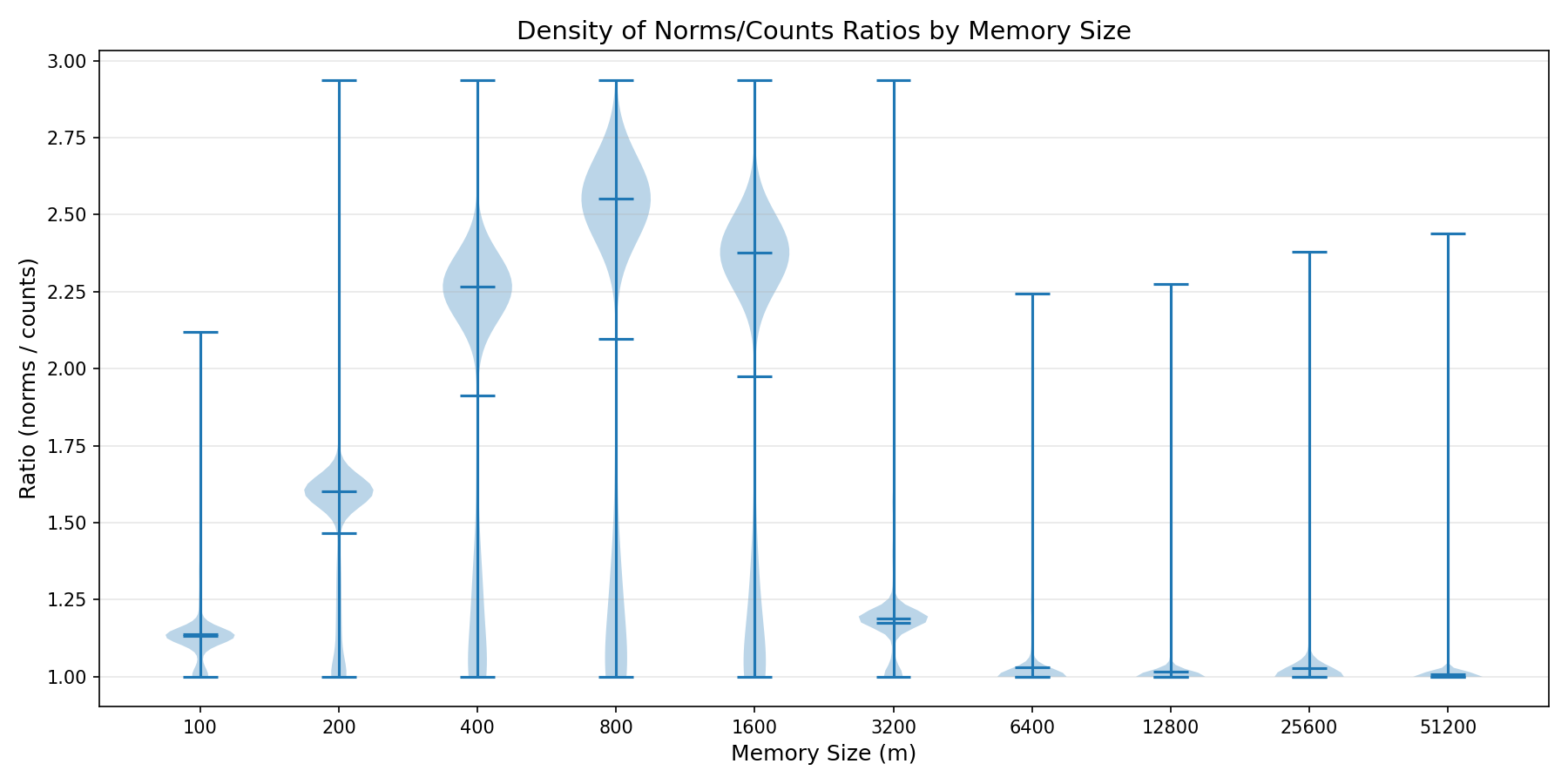}

\caption{Violin plot of the improvement factor from adding $L_2$ and $L_3$ norm inequalities to an entropic bound based only on nonzero counts, for 500 randomly selected matrix pairs from the SuiteSparse collection. Matrices were sliced or truncated where needed to obtain compatible dimensions.}
\end{figure}

Informally, we conjecture that ``structure disappears'' at excessively fine or coarse
resolutions. As a motivating example, consider a ``block staircase'' matrix $A$ where nonzeros
are very dense in the lower-left corner and sparse elsewhere.

In the case where segment sizes are substantially larger than the block
size, the skew makes it necessary for an optimal partitioning to be non-uniform,
with a mix of skinny and near-square segments (as occurred with our example in Figure \ref{fig:pathological-segment-size}). Thus, structural information is only useful when tile sizes are larger than the block
size. 
If, on the other hand, fast memory can store an entire matrix, that matrix can be held stationary, with the other matrices streaming through fast memory, for a total communication cost equal to the size of the input and output matrices. As such, the potential gains
from improved partitioning schemes are negligible, and so structural
factors have little influence on the I/O lower bounds.

One straightforward application of this is automating the evaluation
of tradeoffs when applying preprocessors such as \citet{IXDB25-Improving}
and \citet{VDY05-OSKI}. If the problem
matches the structure-sensitive regime, this suggests the potential
speedup from structure-aware runtime tuning is high;
conversely, in the structure-insensitive regime, the improvements will
be limited since the intensity ceiling is independent of structure.
We leave formalizing this observation
and further exploring its implications for algorithm selection to
future work. 

\subsection{Reduction Count Norm}

The compression ratio is a commonly used measure of the correlation of the sparsity structures of two matrices.
\begin{defn}The \emph{compression ratio} is the ratio of the total number of intermediate products to the number of nonzeros in the output; i.e., the average number of partial sums that generate one nonzero element of the output. \cite{compression-ratio}
\end{defn}

\begin{defn}[Reduction Count Norm]
\label{def:ecr-def}
The $L_p$ reduction count norm, denoted $\gamma_p$, is the $L_p$ norm of the vector $\deg(j \vert ik)$ that describes how many flops $(i,j,k)$ contribute to each output element $C[i,k]$. We also call the number of contributing flops the \textit{reduction length} for $(i,k)$.
\end{defn}

Immediately, we can bound the conditional entropy $h(j \vert ik)$ by the logarithm of the maximum reduction length:
\[h(j \vert ik) \le \lg(\gamma_{\infty}) \]

We can apply \Thmref{lp-norm-bound} to this to extend it and obtain stronger inequalities, similar to the previous section.

\begin{equation}
h(ik) + 2 h(j \vert ik) \le \lg(\gamma_2^2). \label{ecr2norm}
\end{equation}

Following the approach from the previous section, we derive lower bounds by plugging this inequality into the optimization problems of Theorems~\ref{thm:serial-bound} and~\ref{thm:parallel-bound}.

In the parallel case, we obtain the new lower bound:
\[
\frac{\nflops}{\nprocs\gamma_{2}}
\]

As with the $L_2$ and $L_3$ norm constraints, inequalities such as (\ref{ecr2norm}) can strengthen the bounds under the summed memory constraint (\ref{eq:summem}), but the resulting bounds are dominated by input and output size under the relaxation (\ref{eq:conditional-mem-constraints}), which permits symbolic solutions.

\subsection{Spectral norm bounds}
The following spectral norm reflects the adjacency pattern, including information
beyond separate degree moments.
Let $\widehat{A}$ be a $0-1$ matrix with the same sparsity structure
as $A$, and let $\norm{\widehat{A}}_{2}$ be its spectral norm.
\begin{thm}
For every distribution supported on the nonzero entries of $A$, 
\[
h(ij)-\frac{h(i)+h(j)}{2}\le\lg\norm{\widehat{A}}_{2}.
\]
The conditional inequality also holds: 
\[
h(ij\mid s)-\frac{h(i\mid s)+h(j\mid s)}{2}\le\lg\norm{\widehat{A}}_{2}.
\]
\end{thm}

\begin{proof}
Let $p_{ij}$ be any joint distribution supported on the nonzero entries
of $A$, with marginals $p_{i}$ and $p_{j}$. Set $x_{i}=\sqrt{p_{i}}$
and $y_{j}=\sqrt{p_{j}}$, so $\norm x_{2}=\norm y_{2}=1$ and 
\begin{align*}
Z & \coloneqq\sum_{i,j}\widehat{A}[i,j]\sqrt{p_{i}p_{j}}\\
 & =x^{\top}\widehat{A}y\le\norm{\widehat{A}}_{2}.
\end{align*}

Nonnegativity of the KL divergence from $p_{ij}$ to $\widehat{A}[i,j]\sqrt{p_{i}p_{j}}/Z$
gives 
\[
h(ij)-\frac{h(i)+h(j)}{2}\le\lg Z\le\lg\norm{\widehat{A}}_{2}.
\]
Applying this inequality to each conditional distribution given $s$
and averaging over $s$ proves the conditional bound.
\end{proof}

Analogous inequalities hold for $B$ and $C$ with $(i,j)$ replaced
by $(j,k)$ or $(i,k)$ respectively.


\subsection{Linear Functions}

Some matrix structures can be described with simple bounds on the sizes of the images of \emph{linear functions} of the loop indices. For example, if $A$ is a tridiagonal matrix, then $i-j$ can only take on three values.

Linear functions of loop indices have been explored using \emph{discrete Brascamp-Lieb inequalities}, as in \citet{CDK+13-Communication,CDK+15-HolderBrascampLieb}. An entropic formulation of these inequalities can be found in \citet{BCDZ25-Combinatorial}, although enumerating the strongest possible Brascamp-Lieb inequalities is often computationally intractable, requiring approximation algorithms such as that of \citet{OLP+19-Automated}.

Over our datasets, attempts to automatically recognize linear-like structures using Hough transforms and derive constraints from them did not result in stronger bounds than those provided by previous inequalities; these bounds are therefore omitted from our experiments. It is possible, however, that these bounds may be of use for highly specific sparsity structures (e.g. banded), and we leave exploration of this to future work.

\section{Empirical Analysis}

\label{sec:empirical}

To test the practicality of our bounds, we assess whether sparsity
information gives useful estimates of peak communication efficiency.
We compare measured communication with entropic bounds, an Erdős--Rényi-based estimate assuming uncorrelated inputs, and previous theoretical baselines: Pagh--Stöckel for the serial case and Greiner for the parallel case.

We note that these bounds do not measure the same thing: Pagh--Stöckel
and Greiner are communication lower bounds on the \emph{worst }matrices
of the same size as the input matrices (and therefore can - and do
- generate figures sufficiently pessimistic that they can be beaten
by actual implementations), and Erdős--Rényi assumes structure that
most matrices do not have. Furthermore, we do not expect measured
performance to be extremely close to our predicted model due to library
overheads, inefficient implementations, etc. Our goal here is to see
how well our bounds compare with other theoretical models of SpGEMM
cost in terms of \emph{correlation} with real-world performance.

\subsection{Experimental setup}

All experiments were performed on an Intel Xeon 6246R (Cascade Lake)
system, using one core for serial experiments and all 16 physical
cores (with hyper-threading disabled) on one socket for parallel experiments.
Each core has a private, 16-way set-associative, 1 MiB L2 cache with
64 byte lines. As both metadata indices and values were 64-bit values, we used $M=1\text{MiB}/(2\times 8\text{B})=65536$. All three references use this same two-word
conversion.

The implementations we used were Intel oneMKL 2025.2.0, SuiteSparse
GraphBLAS 7.2.0 \citep{Dav19-Algorithm}, and (for the parallel case)
the released implementation of Islam et al. \citep{IXDB25-Improving},
which incorporates a wide variety of preprocessing steps (including
permutation and partitioning). As we are interested in \emph{peak}
performance, we try to call these libraries in a way that maximizes
their performance, and to benchmark only the numeric computation,
without measuring the cost of preprocessing, format construction,
correctness checking, or symbolic phases (except for GraphBLAS), and
do not flush the cache between symbolic and numeric phases.

For MKL, we measure only the numeric \texttt{FINALIZE\_MULT} phase,
with a fresh symbolic \texttt{NNZ\_COUNT} phase performed before each
measurement. For Islam et al., we similarly exclude symbolic/hash
preparation and retain its native output format. As GraphBLAS does
not expose stages to users, we measured the complete multiplication,
including completion of deferred work. 

Based on experiments on the fastest layouts for our machine and datasets,
we used CSR inputs and output for MKL, and CSC for GraphBLAS. For
Islam et al.'s implementation, we retained each kernel’s native CSR
or clustered-CSR representation and selected kernel with the lowest
measured L2 traffic. We report medians of three repetitions after
one warmup. 

\subsubsection{Dataset}

For our main experiments, following the methodology of \citet{IXDB25-Improving},
we compute squares $A^{2}$ of real SuiteSparse matrices \citep{KAB+19-SuiteSparse}.\footnote{The square-product focus supports our graph-partitioning experiments.
As a supplementary check, we formed 13 products from SuiteSparse matrices
in matching application categories whose operand patterns are neither
equal nor transposes; two have rectangular operands. Under original-order
MKL \texttt{FINALIZE\_MULT}, measured incoming traffic exceeds the
degree/reduction count norm estimate with spectral bounds on all 13
products, with a median ratio of $2.70$ (11 within a factor of four).} Our symmetric cohort selects structurally symmetric inputs with $2^{17}$
to $2^{27}$ nonzeros and at most ten representatives per collection
group; eleven complex-valued inputs are excluded. The serial experiments
contain 195 products with at most $10^{8}$ structural scalar products.
For each input, we retain the lowest incoming traffic observed for
MKL over original and nested-dissection orders, supplemented by a
search over random and METIS permutations. 

The parallel experiments contain 163 symmetric-cohort products with
$\nflops\le16\cdot10^{9}$, after excluding overly small inputs requiring
fewer than five estimated segments per core. We also evaluate 68 real
square inputs from the corpus of Islam et al., using the same thresholds;
29 overlap the symmetric cohort. We ran both the original order and
the hierarchical clustering order, forming $(PA)(AP^{\mathsf{T}})=PA^{2}P^{\mathsf{T}}$
(where $P$ is a permutation matrix), and selecting each implementation's
configuration using minimum median aggregate incoming traffic. One
input in the Islam corpus has only original-order results because
permutation preparation exceeded its memory allocation.

\subsubsection{Measurements}

For read traffic, we collect \texttt{L2\_LINES\_IN:ALL}, using PAPI
\citep{JDC+25-Advancements} for serial runs and inherited, CPU-filtered
Linux performance events for parallel runs. Multiplying the count
by eight gives incoming traffic in 8-byte words. We also collect the
number of writebacks \texttt{L2\_TRANS:L2\_WB}, which, when summed
with the read traffic count, gives a fills-plus-writeback proxy. Our
primary figures use incoming traffic, which is a reasonable approximation
of total traffic as each write (i.e. accumulate) must be accompanied
by a read of the existing partial sum (the graphs for the fills-plus-writeback proxy look extremely similar). As in Section 2, $\nflops$
counts structural scalar products, with one multiply-add counted as
one unit.

\subsection{Serial bounds}

The entropic calculation includes support sizes, row/column degree/norms
of orders $1,2,3,\infty$, and the spectral inequalities. 

We computed the entropic bound using
numerical optimization, which allowed us to constrain the sum of the
memory footprints of each segment to fit in cache (rather than imposing
the separate per-matrix constraints of (\ref{eq:conditional-mem-constraints})).

Writing $N=\nnz A+\nnz B$ and $Z=\nnz C$, we used
\begin{align*}
Q_{\mathrm{PS}} & =2N\min\!\left\{ \sqrt{Z/M},N/M\right\} 
\end{align*}
for the Pagh--Stöckel bound and
\[
Q_{\mathrm{ER}}=2\sqrt{\nnz A\,\nnz B\,\nnz C/M}
\]
for a bound that assumes uncorrelated Erdős--Rényi inputs and outputs.

\begin{figure*}[t]
 \includegraphics[width=1\textwidth]{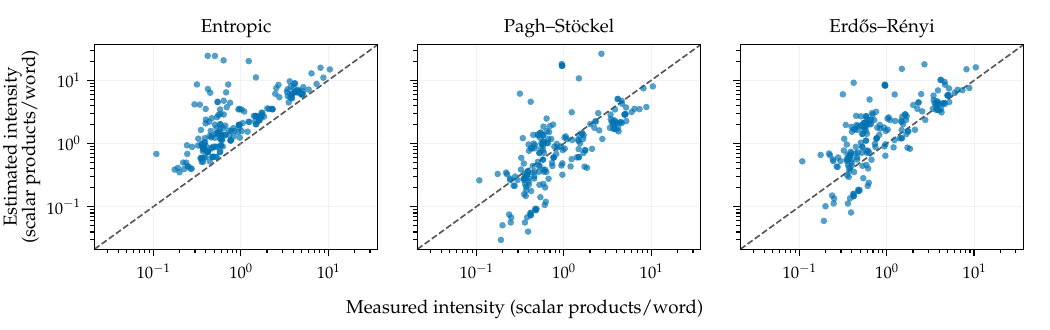}
\caption{Serial MKL numeric-phase intensity versus three raw estimates; dashed
lines indicate equality. Work counts structural scalar products, and
traffic counts 8-byte words, with two words per modeled entry. The entropic bound
includes spectral and ECR constraints, but no segment-work bound.}
\Description{Three log-log scatter plots compare measured serial
intensity with entropic, Pagh--Stöckel, and Erdős--Rényi estimates.}
\label{fig:empirical-serial} 
\end{figure*}

\begin{figure*}[t]
 \includegraphics[width=1\textwidth]{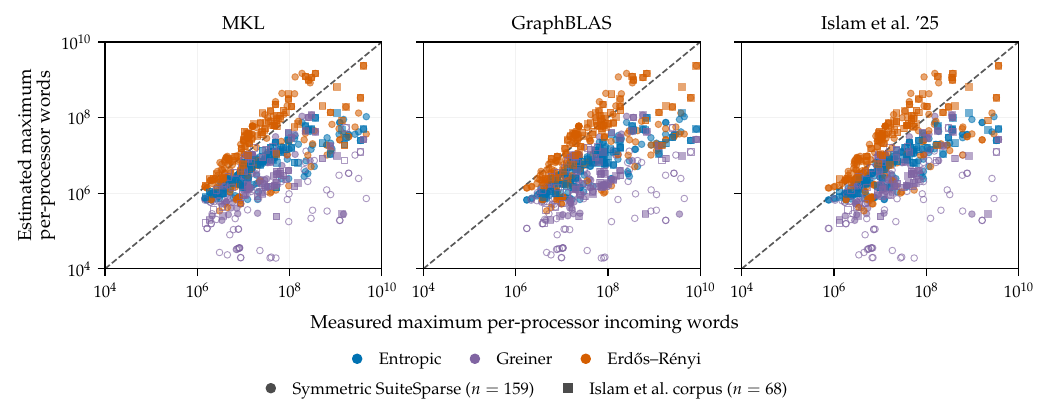}
\caption{Maximum per-processor incoming communication versus estimates (8-byte
words), with shared log scales and dashed equality lines. All three estimates
use the same computable products. Hollow Greiner markers indicate
parameter-range extrapolations; no points are clipped.}
\Description{Three log-log scatter plots compare measured maximum
per-processor incoming communication with entropic, Greiner, and Erdős--Rényi
estimates for MKL, GraphBLAS, and Islam et al. 2025. Circles identify
symmetric SuiteSparse inputs and squares identify the Islam corpus.}
\label{fig:empirical-parallel} 
\end{figure*}

Figure \ref{fig:empirical-serial} compares measured arithmetic intensity
with $\nflops/Q$. The entropic estimate exceeds measured intensity
on all 195 products, with a median ratio of $3.36$ and an interquartile
range of $2.42$--$7.91$; 115 products are within a factor of four.
The Pagh--Stöckel and Erdős--Rényi estimates, being worst-case bounds,
fall below measured intensity on 128 and 52 products, respectively
(i.e. far too pessimistic).

\subsection{Parallel bounds}

For the parallel case, we minimize $h(s\mid p)$ with $h(ijk)=\lg\nflops$
and $h(p)=\lg16$, and use $2M2^{\min h(s\mid p)}$ as the maximum
per-processor communication estimate. We retain the same degree norm
and reduction count norm inequalities as in the serial comparison
(spectral norm constraints did not appear to strengthen the bounds
in this case).

For the parallel Erd\H{o}s--R\'enyi comparison, we divide the serial
reference by $P=16$, retaining the per-processor cache capacity $M$:
\[
\frac{Q_{\mathrm{ER}}}{P}
=
\frac{2}{P}
\sqrt{\frac{\nnz A\,\nnz B\,\nnz C}{M}}.
\]

Greiner's per-processor progress argument (Lemma 2.4 and Theorem 6.2
of \citep{Gre12-Sparse}) gives a bound on maximum-processor communication.
For square dimension $n$, $k_{1}=\nnz A/n$, $k_{2}=\nnz B/n$, and
$D$ from his Eq. (6.1), we use $2k_{1}k_{2}n/[16\min\{D,\sqrt{M}\}]$,
with $\epsilon=0.1$ and suppressed asymptotic constants. The theorem
is a worst-case bound and assumes ``pseudo-rectangular algorithms'',
which the libraries we use do not necessarily qualify as (although
it conjectures that these bounds are universally applicable). The
expression is computable for 159/163 symmetric and 68/68 Islam-corpus
products; its parameter range holds for 56 and 48, respectively. We
retain extrapolations explicitly as reference values, not as instance-specific
bounds.

Figure \ref{fig:empirical-parallel} shows that the entropic estimates
are generally closer to the measured communication than Greiner's
reference, and cluster more tightly. For the symmetric cohort, the
median measured/estimated ratios are $3.43$, $6.07$, and $2.71$
for MKL, GraphBLAS, and Islam et al., respectively, compared with
$10.32$, $19.82$, and $8.45$ for Greiner. For the Islam corpus,
the corresponding medians are $3.80$, $7.00$, and $3.04$, compared
with $11.57$, $21.96$, and $8.91$. The same ordering of median
gaps holds when restricted to Greiner's parameter range.

The Erdős--Rényi estimate has a smaller median gap from measured communication
than either theoretical bound: the measured/estimated ratios are $1.07$,
$1.88$, and $0.83$ for MKL, GraphBLAS, and Islam et al., respectively,
on the symmetric cohort, and $0.61$, $1.06$, and $0.52$ on the Islam corpus.
However, its estimates are much more widely scattered than the entropic
estimates, reducing its value as a predictor.

Table~\ref{tab:parallel-correlations} reports the Pearson correlations
between measured communication and each estimate. Greiner's correlation with the measured cost is extremely poor, and our entropic bound solidly outperforms Erdős--Rényi as a predictor of real performance.

\begin{table*}[t]
\caption{Pearson correlations between measured and estimated maximum per-processor communication on the raw (untransformed) scale, with the symmetric SuiteSparse and Islam et al. cohorts combined.}
\label{tab:parallel-correlations}
\centering
\begin{tabular}{llccc}
\hline
Measured metric & Implementation & Entropic bound & Greiner & Erd\H{o}s--R\'enyi \\
\hline
Incoming fills & MKL & \textbf{0.604} & 0.115 & 0.312 \\
Incoming fills & GraphBLAS & \textbf{0.627} & 0.138 & 0.423 \\
Incoming fills & Islam et al. & \textbf{0.602} & 0.105 & 0.382 \\
Fills + writebacks & MKL & \textbf{0.602} & 0.142 & 0.381 \\
Fills + writebacks & GraphBLAS & \textbf{0.610} & 0.159 & 0.481 \\
Fills + writebacks & Islam et al. & \textbf{0.594} & 0.111 & 0.390 \\
\hline
\end{tabular}
\end{table*}

\section{Discussion, Conclusion, and Future Work}

This paper has described an information-theoretic framework that allows
instance-specific communication lower bounds to be algorithmically
constructed. Our experiments have shown that the resulting bounds
can (to an extent greater than previous bounds) serve a role as a
performance predictor and optimization target for hardware and software
SpGEMM implementations. The process to generate these bounds can immediately
be applied to higher-order sparse tensor contractions as well, although
we omit such an analysis for lack of space and a baseline model with
which we can compare.

An immediate question that may be raised is whether this work can
be used to develop communication-avoiding algorithms, just as dense
communication lower bounds often lead to algorithms that achieve those
bounds; we have had some success experimenting with developing communication-optimal
\emph{dense} parallel algorithms for tensor contraction by generating
distributions under which the entropic optimization program is tight
(i.e., automatically generating ``N.5D'' algorithms \citet{SD11-CommunicationOptimal}).
While obtaining an instance-specific \emph{optimal} algorithm for
SpGEMM is infeasible and likely intractable, dualizing the LP can
provide an interesting starting point for searching for good algorithms.
Alternatively, our bounds can be used as cheap cost models for load
balancing or algorithm selection.

Our framework is extensible, and we expect that future work will be
able to strengthen the model by analyzing constant factors more carefully
and incorporating other matrix properties and sparsity structures,
or to bring it closer to the hardware, e.g., by supporting distributed-memory
machines and sparse hardware accelerators, or by modeling metadata
costs.

We also believe that our methods can be used to evaluate \emph{sparsity
structures}. For applications where sparsity is a variable under the
user's control (e.g., designing sparse attention mechanisms), a communication
cost model can provide early feedback.

An additional application that has shown some promise in early experiments
is using this lower bound framework as part of an agentic loop. Given
a small set of input matrices with some common structure (e.g., from
a practical workload), an agent can be prompted repeatedly to optimize
an implementation and tighten the lower bound toward the implementation's
communication cost. In early experiments, this combined approach appears
more effective than prompting the agent to work on either task alone.
We leave this experimentation to future work.

\section{Acknowledgements}

The authors would like to thank Mahmoud Abo Khamis, Kyle Deeds, Hengrui Luo, and Dan Suciu, for providing trenchant feedback, comments, and discussions. 

Grace Dinh's research was supported in part by ADIA Lab, Fall 2025.

AI agents, primarily ChatGPT (5.6 Sol and 6 Astra) and Claude (Opus
and Fable), were used to assist with coding the interface to MPT3
used to symbolically solve for the lower bounds, the cache benchmarks,
and the code used to generate the plots in Section \ref{sec:empirical}.
We also used these tools for proofreading/editing, literature search,
 proof checking, and algebraic grunt work.

\bibliographystyle{ACM-Reference-Format}
\addcontentsline{toc}{chapter}{\bibname}\bibliography{gnd,overleaf-additions}

\end{document}